\documentclass[osajnl,twocolumn,showpacs,showkeys,10pt]{revtex4}  

\usepackage{graphicx}
\usepackage{dcolumn}
\usepackage{bm}

\newcommand{\ud}{\mathrm{d}}
\newcommand{\eqn}[1]{(\ref{#1})}
\newcommand{\fig}[1]{Fig. \ref{#1}}

\begin{document}
%
\title{Maximum-Likelihood Detection of Soliton with Timing Jitter}
\author{Keang-Po Ho}
\affiliation{Institute of Communications Engineering and Department of Electrical Engineering, National Taiwan University, Taipei 106, Taiwan.}
\email{kpho@cc.ee.ntu.edu.tw}

\date{\today}%
%
\begin{abstract}
Using the maximum-likelihood detector (MLD) of a soliton with timing jitter and noise, other than walk-out of the bit interval, timing jitter does not degrade the performance of MLD.
When the MLD is simulated with important sampling method, even with a timing jitter standard deviation the same as the full-width-half-maximum (FWHM) of the soliton, the signal-to-noise (SNR) penalty is just about 0.2 dB.
The MLD performs better than conventional scheme to lengthen the decision window with additive noise proportional to the window wide. 
\end{abstract}


\maketitle

The Gordon-Haus timing jitter \cite{gordon86} limits the transmission distance of a soliton communication system.
The arrival time of the soliton has a variance increase cubically with distance.
Previously, the decision window of the soliton is widened to significantly reduce the impact of timing jitter \cite{iannone}.
However, the widening of the decision window allows more noise entering the decision circuits.
For example, if the decision window is doubled to twice wider than necessary, the amount of noise is doubled. 
The signal-to-noise ratio (SNR) is halved, giving 3-dB SNR penalty to the system. 
Using an electro-absorption modulator as an optical time-domain demultiplexer to provide a wide decision window \cite{suzuki92}, the timing window may reach 80\% of the bit interval for timing-jitter resilient reception  \cite{mollenauer96}.

In conventional detection theory \cite{mcdonough2}, the matched filter based receiver maximizes the output SNR.
The matched filter can be implemented optically with an impulse response identical to the soliton pulse shape.
Unfortunately, the matched filter cannot apply directly to a signal with timing jitter.
A wide decision window is equivalent to an intergator with an integration interval the same as the decision window.
Even for a soliton with timing jitter, a rectangular decision window is inferior to the match filter based receiver.
Method to combat timing jitter without leading to significant increase in SNR penalty is investigated here based on maximum-likelihood detection (MLD).

MLD of a signal minimizes the error probability of the detection of a binary signal.
If digital ``1'' and ``0'' are represented by the presence or absence of a soliton and assumed that ``1'' or ``0'' is transmitted with equal probability, MLD decides the presence of a soliton by $p[r(t)|1] > p[r(t)|0]$, where $r(t)$ is the received signal, $p[r(t)|1]$ and $p[r(t)|0]$ are the probability of having a received signal of $r(t)$ given the condition with the presence or absence of a soliton, respectively.
The absence of a soliton is decided if $p[r(t)|1] < p[r(t)|0]$.

In a soliton communication system, the received signal can be represented as 

\begin{equation}
 r(t) = a_k s(t-t_0)e^{j \phi} + n(t),
\label{solrt}
\end{equation}

\noindent where $a_k \in \{0, 1\}$ for the absence or presence of the soliton, $s(t) = \mathrm{sech}(1.76t)$ is the normalized soliton pulse with unity full-width-half-maximum (FWHM), $t_0$ is a random variable representing the timing jitter, $\phi$ is the random phase due to the propagation delay and soliton phase jitter, and $n(t)$ is the additive complex-value white Gaussian noise with spectral density of $N_0/2$.
Only the noise with the same polarization as the soliton is considered here by assuming a polarized receiver.
The phase of $\phi$ is assumed to be uniformly distributed from 0 to $2 \pi$.

Usually, soliton propagation with noise is studied by the first-order perturbation of the soliton \cite{kivshar89, kaup90, georges95, iannone} in which amplifier noise is directly projected to amplitude and frequency jitter.
When the first-order soliton perturbation is linearized \cite{iannone}, there is no difference whether amplitude jitter is a distributed contribution along the fiber or a lumped contribution at the beginning or the end of the fiber.
For example, if $n(t) = n_1(t) + n_2(t)$ with $n_1(t)$ and $n_2(t)$ from the first and second half of the fiber link, respectively.
With the small signal or linearized model \cite{iannone}, the projection of $n(t)$ to amplitude and frequency jitter is the same as first applied $n_1(t)$ and then $n_2(t)$, or even first applied $n_2(t)$ and then $n_1(t)$.
Of course, if first-order large signal perturbation is used, there is small difference between the distributed or lumped model \cite{iannone, ho0311, moore03}.
The received signal of \eqn{solrt} assumes all amplifier noise at the end of the fiber link and is more accurate than first-order perturbation for amplitude and frequency jitter if time jitter is included in $t_0$ and phase jitter is included in $\phi$.

If the soliton is detected by a photodetector, the phase of $\phi$ in \eqn{solrt} does not affect the system performance.
With a detail provided in [\onlinecite{mcdonough2}, Sec. 7.2], after averaging over the random phase of $\phi$, the probability density of the received signal is equal to

\begin{eqnarray}
p[r(t)|1, t_0] &=& \alpha \exp\left( - \frac{1}{N_0} \int_{-\infty}^\infty |r(t)|^2 \ud t - \frac{E}{N_0} \right) \nonumber \\
     & &   \times    I_0 \left( \frac{2\sqrt{E}q} {N_0}\right), 
   \label{soldetpr1t0}\\
p[r(t)|0] & = & \alpha \exp\left( - \frac{1}{N_0} \int_{-\infty}^\infty  |r(t)|^2 \ud t \right).
   \label{soldetpr0}
\end{eqnarray}

\noindent where $\alpha$ is a proportional constant,  $I_0(\ )$ is the zero-order modified Bessel function of the first kind, $E = \int_{-\infty}^{\infty} s^2(t) \ud t $ is the energy per soliton pulse, and $q$ is equal to 

\begin{equation}
q = \left| \int_{-\infty}^{\infty} r(t)s(t-t_0) \ud t \right|,  \quad q \geq 0.
\label{soldetq}
\end{equation}

If the probability density of timing jitter is $p_T(t_0)$, we obtain 

\begin{equation}
p[r(t)|1] = \int_{-\infty}^{\infty} p[r(t)|1, t_0] p_T(t_0) \ud t_0.
\end{equation}

\noindent Using the likelihood ratio of $p[r(t)|1]/p[r(t)|0]$, the decision rule  becomes

\begin{equation}
\int_{-\infty}^{\infty} I_0 \left(  \frac{2 \sqrt{E} q} {N_0}  \right)  p_T(t_0) \ud t_0 
{{1 \atop >} \atop {< \atop 0}}   \exp\left( \frac{E}{N_0} \right)
\label{decrule}
\end{equation}

\noindent for the presence or absence of a soliton with time jitter.
In the decision rule of \eqn{decrule}, the integration of $|r(t)|^2$ in \eqn{soldetpr1t0} and \eqn{soldetpr0} and the constant of $\alpha$ cancel each other.

\begin{figure}
\centerline{
  \includegraphics[width = 0.47 \textwidth]{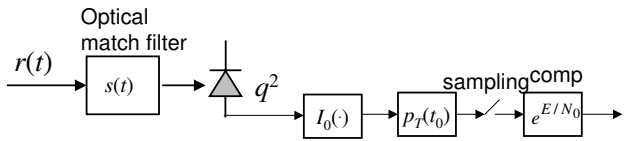}
}
\caption{MLD of the presence and absence of a soliton with timing jitter.}
\label{figdet}
\end{figure}

The decision rule of \eqn{decrule} together with the parameter $q$ calculated by \eqn{soldetq} can be implemented by the block diagram of Fig. \ref{figdet}.
The received signal first passes thought an optical matched filter having an impulse response equal to the soliton pulse of $s(t)$.
The output of the optical matched filter is $qe^{j \phi}$.
The output of the optical matched filter converts to electrical signal using a photodetector.
The photodetector gives an output proportional to the square of $q^2$.
The implementation of the correlation of \eqn{soldetq} using matched filter can be found, for example, in [\onlinecite{mcdonough2}, ch. 6].
With the output of $q^2$ from the photodetector, the value of $I_0(2 \sqrt{E} q/N_0)$ in \eqn{decrule} can be found.
The integration in \eqn{decrule} with respect to $t_0$ is again implemented using a filter with impulse response of $p_T(t_0)$ which output is sampled at the right time.
In \fig{figdet}, the probability density of $p_T(t_0)$ is not necessary to be Gaussian distributed \cite{menyuk95, ho0311} but must be symmetrical with respect to zero.
After the sampler, the presence or absence of the soliton is decided when compared with $\exp(E/N_0)$.

\begin{figure}
\centerline{
  \includegraphics[width = 0.35 \textwidth]{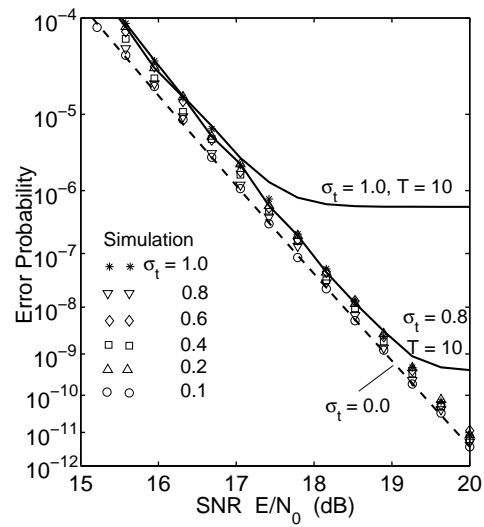}
}
\caption{Simulated error probability of the MLD for soliton with timing jitter.
Various markers are the error probability from simulation.
Dashed-line is the theoretical error probability without timing jitter.
Solid lines include the walk-out probability that the timing jitter is outside the bit interval.
}
\label{figber}
\end{figure}

Without timing jitter or $p_T(t_0) = \delta(t_0)$, the right-hand sided of \eqn{decrule} becomes $I_0(2 \sqrt{E} q/N_0)|_{t_0 = 0}$ and the decision rule of \eqn{decrule} can be simplified to a quadratic detector [\onlinecite{mcdonough2}, Sec. 8.3].
The quadratic detector is $q^2 {{1 \atop >} \atop {< \atop 0}} q^2_\mathrm{th}$ with $q_\mathrm{th}$ as the optimal threshold without timing jitter.
With a performance the same as that for non-coherent detection of amplitude-shift keying signal, the performance can be analyzed by the well-known Marcum $Q$-function \cite{marcum60, yamamoto80}.
The error probability for the case without timing jitter is shown in \fig{figber} as dashed line.
The error probability of \fig{figber} is shown as a function of SNR, given by the ratio of $E/N_0$.
The threshold of detection is calculated using \eqn{decrule} with $p_T(t_0) = \delta(t_0)$.
An error probability of $10^{-9}$ requires an SNR about $18.9$ dB.

The performance of the MLD of \eqn{decrule} does not lead to a simple analytical error probability for soliton with timing jitter.
Numerical simulation is conducted if the timing jitter is zero-mean Gaussian distributed with variance of $\sigma_t^2$.
The simulation results are shown in \fig{figber} with $\sigma_t$ normalized to the FWHM of the soliton.
\fig{figber} shows that MLD for soliton with timing jitter has very small SNR penalty for $\sigma_t$ up to one FWHM of the soliton. 

In order to investigate those cases with very small error probabilities, numerical simulation cannot be conducted directly based on Monte-Carlo methods.
The simulation of \fig{figber} is based on important sampling, similar to the methods of [\onlinecite{shanmugam80, moore03}].
The received signal has soliton with different timing jitter according the Gaussian distribution with variance of $\sigma_t^2$.
The noise sample after the optical matched filter of \fig{figdet} with a time corresponding to the peak optical intensity is generated based on uniform distribution.
Other noise samples are generated by Gaussian distribution with a covariance depending on the optical matched filter.
Each error count is weighted according to the probability difference between the actual Gaussian noise samples with the generated noise samples \cite{moore03, shanmugam80}.
Other than adding a biased noise sample after the optical matched filter of \fig{figdet} instead of the actual signal with amplifier noises before the filter, the numerical simulation followed closely the detector of \fig{figdet}.

The error probability calculated from simulation is shown in \fig{figber} using different marker for $\sigma_t$ from $0.1$ to $1.0$ of the FWHM of the soliton.
Even with a soliton having a large timing jitter of $\sigma_t = 1.0$, the SNR penalty is just about 0.2 dB compared with the case without timing jitter (dashed line).
If a widening decision window is used instead, for example, $\sigma_t = 0.5$ requires a decision window wide of $\tau_w \approx 6$ such that the probability of the soliton walking-out of the decision window is less than $\mathrm{erfc}\!\!\left[\tau_w/(2\sigma_t\sqrt{2})\right] = 2 \times 10^{-9}$ [\onlinecite{gordon86}].
A decision window size of $\tau_w \approx 6$ gives approximately 7 dB of SNR penalty.

\fig{figber} also shows the error probability taking into account the probability that the soliton may have a timing jitter outside the bit interval when $\sigma_t = 0.8$, $1.0$ and the bit interval is 10 times the FWHM of the soliton.
The bit interval of $T = 10$ is chosen for convenience \cite{gordon86}. 
For $\sigma_t < 0.8$, the walk-out probability does not affect the overall error probability and is not shown in \fig{figber}.
From \fig{figber}, soliton with large timing jitter is mainly affected by the walk-out probability, especially for system with a bit interval just $T =4, 6$ times the FWHM of the soliton.
Unlike the receiver with widening decision window, the simulation results of \fig{figber} show that the receiver schematic of \fig{figdet} does not give large SNR penalty.

The MLD of \eqn{decrule} or \fig{figdet} has a filter with impulse response the same as the probability density of $p_T(t_0)$.
The walk-out probability depends on the tail of $p_T(t_0)$ but the left hand-sided of \eqn{decrule} depends on the center of $p_T(t_0)$ around its mean of $t_0 = 0$.
While the MLD of \eqn{decrule} depends weakly on the timing jitter variance of $\sigma_t^2$, the walk-out probability depends strongly on $\sigma_t^2$ as from \fig{figber}.   

The MLD of soliton with timing jitter is derived, to our knowledge, the first time.
Other than the walk-out probability that the soliton has a timing jitter outside the bit-interval, soliton is not affected by timing jitter when MLD is used.
Even with a timing jitter standard deviation the same as the soliton FWHM, the SNR penalty is just about 0.2 dB.
The MLD has significantly smaller SNR penalty than detector with a widening decision window.


\begin{thebibliography}{14}
\expandafter\ifx\csname natexlab\endcsname\relax\def\natexlab#1{#1}\fi
\expandafter\ifx\csname bibnamefont\endcsname\relax
  \def\bibnamefont#1{#1}\fi
\expandafter\ifx\csname bibfnamefont\endcsname\relax
  \def\bibfnamefont#1{#1}\fi
\expandafter\ifx\csname citenamefont\endcsname\relax
  \def\citenamefont#1{#1}\fi
\expandafter\ifx\csname url\endcsname\relax
  \def\url#1{\texttt{#1}}\fi
\expandafter\ifx\csname urlprefix\endcsname\relax\def\urlprefix{URL }\fi
\providecommand{\bibinfo}[2]{#2}
\providecommand{\eprint}[2][]{\url{#2}}

\bibitem[{\citenamefont{Gordon and Haus}(1986)}]{gordon86}
\bibinfo{author}{\bibfnamefont{J.~P.} \bibnamefont{Gordon}} \bibnamefont{and}
  \bibinfo{author}{\bibfnamefont{H.~A.} \bibnamefont{Haus}},
  \bibinfo{journal}{Opt. Lett.} \textbf{\bibinfo{volume}{11}},
  \bibinfo{pages}{865} (\bibinfo{year}{1986}).

\bibitem[{\citenamefont{Iannone et~al.}(1998)\citenamefont{Iannone, Matera,
  Mecozzi, and Settembre}}]{iannone}
\bibinfo{author}{\bibfnamefont{E.}~\bibnamefont{Iannone}},
  \bibinfo{author}{\bibfnamefont{F.}~\bibnamefont{Matera}},
  \bibinfo{author}{\bibfnamefont{A.}~\bibnamefont{Mecozzi}}, \bibnamefont{and}
  \bibinfo{author}{\bibfnamefont{M.}~\bibnamefont{Settembre}},
  \emph{\bibinfo{title}{Nonlinear Optical Communication Networks}}
  (\bibinfo{publisher}{John Wiley {\&} Sons}, \bibinfo{address}{New York},
  \bibinfo{year}{1998}).

\bibitem[{\citenamefont{Suzuki et~al.}(1992)\citenamefont{Suzuki, Tanaka,
  Edagawa, and Matsushima}}]{suzuki92}
\bibinfo{author}{\bibfnamefont{M.}~\bibnamefont{Suzuki}},
  \bibinfo{author}{\bibfnamefont{H.}~\bibnamefont{Tanaka}},
  \bibinfo{author}{\bibfnamefont{N.}~\bibnamefont{Edagawa}}, \bibnamefont{and}
  \bibinfo{author}{\bibfnamefont{Y.}~\bibnamefont{Matsushima}},
  \bibinfo{journal}{J. Lightwave Technol.} \textbf{\bibinfo{volume}{10}},
  \bibinfo{pages}{1912} (\bibinfo{year}{1992}).

\bibitem[{\citenamefont{Mollenauer et~al.}(1996)\citenamefont{Mollenauer,
  Mamyshev, and Neubelt}}]{mollenauer96}
\bibinfo{author}{\bibfnamefont{L.~F.} \bibnamefont{Mollenauer}},
  \bibinfo{author}{\bibfnamefont{P.~V.} \bibnamefont{Mamyshev}},
  \bibnamefont{and} \bibinfo{author}{\bibfnamefont{M.~J.}
  \bibnamefont{Neubelt}}, \bibinfo{journal}{Electron. Lett.}
  \textbf{\bibinfo{volume}{32}}, \bibinfo{pages}{471} (\bibinfo{year}{1996}).

\bibitem[{\citenamefont{McDonough and Whalen}(1995)}]{mcdonough2}
\bibinfo{author}{\bibfnamefont{R.~N.} \bibnamefont{McDonough}}
  \bibnamefont{and} \bibinfo{author}{\bibfnamefont{A.~D.}
  \bibnamefont{Whalen}}, \emph{\bibinfo{title}{Detection of Signals in Noise}}
  (\bibinfo{publisher}{Academic Press}, \bibinfo{address}{San Diego},
  \bibinfo{year}{1995}), \bibinfo{edition}{2nd} ed.

\bibitem[{\citenamefont{Kivshar and Malomed}(1989)}]{kivshar89}
\bibinfo{author}{\bibfnamefont{Y.~S.} \bibnamefont{Kivshar}} \bibnamefont{and}
  \bibinfo{author}{\bibfnamefont{B.~A.} \bibnamefont{Malomed}},
  \bibinfo{journal}{Rev. Mod. Phys.} \textbf{\bibinfo{volume}{61}},
  \bibinfo{pages}{763} (\bibinfo{year}{1989}), \bibinfo{note}{addendum:
  \textbf{\bibinfo{volume}{63}}, 211 (1993)}.

\bibitem[{\citenamefont{Kaup}(1990)}]{kaup90}
\bibinfo{author}{\bibfnamefont{D.~J.} \bibnamefont{Kaup}},
  \bibinfo{journal}{Phys. Rev. A} \textbf{\bibinfo{volume}{42}},
  \bibinfo{pages}{5689} (\bibinfo{year}{1990}).

\bibitem[{\citenamefont{Georges}(1995)}]{georges95}
\bibinfo{author}{\bibfnamefont{T.}~\bibnamefont{Georges}},
  \bibinfo{journal}{Opt. Fiber Technol.} \textbf{\bibinfo{volume}{1}},
  \bibinfo{pages}{97} (\bibinfo{year}{1995}).

\bibitem[{\citenamefont{Ho}(2003)}]{ho0311}
\bibinfo{author}{\bibfnamefont{K.-P.} \bibnamefont{Ho}}, \bibinfo{journal}{Opt.
  Lett.} \textbf{\bibinfo{volume}{28}}, \bibinfo{pages}{2165}
  (\bibinfo{year}{2003}).

\bibitem[{\citenamefont{Moore et~al.}(2003)\citenamefont{Moore, Biondini, and
  Kath}}]{moore03}
\bibinfo{author}{\bibfnamefont{R.~O.} \bibnamefont{Moore}},
  \bibinfo{author}{\bibfnamefont{G.}~\bibnamefont{Biondini}}, \bibnamefont{and}
  \bibinfo{author}{\bibfnamefont{W.~L.} \bibnamefont{Kath}},
  \bibinfo{journal}{Opt. Lett.} \textbf{\bibinfo{volume}{28}},
  \bibinfo{pages}{105} (\bibinfo{year}{2003}).

\bibitem[{\citenamefont{Menyuk}(1995)}]{menyuk95}
\bibinfo{author}{\bibfnamefont{C.~R.} \bibnamefont{Menyuk}},
  \bibinfo{journal}{Opt. Lett.} \textbf{\bibinfo{volume}{20}},
  \bibinfo{pages}{285} (\bibinfo{year}{1995}).

\bibitem[{\citenamefont{Marcum}(1960)}]{marcum60}
\bibinfo{author}{\bibfnamefont{J.~I.} \bibnamefont{Marcum}},
  \bibinfo{journal}{IRE Trans. Info. Theory} \textbf{\bibinfo{volume}{IT-6}},
  \bibinfo{pages}{56} (\bibinfo{year}{1960}).

\bibitem[{\citenamefont{Yamamoto}(1980)}]{yamamoto80}
\bibinfo{author}{\bibfnamefont{Y.}~\bibnamefont{Yamamoto}},
  \bibinfo{journal}{IEEE J. Quantum Electron.}
  \textbf{\bibinfo{volume}{QE-16}}, \bibinfo{pages}{1251}
  (\bibinfo{year}{1980}).

\bibitem[{\citenamefont{Shanmugam and Balaban}(1908)}]{shanmugam80}
\bibinfo{author}{\bibfnamefont{K.~S.} \bibnamefont{Shanmugam}}
  \bibnamefont{and} \bibinfo{author}{\bibfnamefont{P.}~\bibnamefont{Balaban}},
  \bibinfo{journal}{IEEE Trans. Commun.} \textbf{\bibinfo{volume}{COM-28}},
  \bibinfo{pages}{1916} (\bibinfo{year}{1980}).

\end{thebibliography}

\end{document}